\documentstyle{article}%
\begin{document}

\title{Protein secondary structure prediction by combining hidden Markov models
and sliding window scores}
\author{Wei-Mou Zheng\\
{\it Institute of Theoretical Physics, Academia Sinica, Beijing
100080, China, }\\ and\\ {\it Bejing Genomics Institute, Academia
Sinica, Beijing 101300, China } } \date{}\maketitle
\begin{abstract}
Instead of conformation states of single residues, refined
conformation states of quintuplets are proposed to reflect
conformation correlation. Simple hidden Markov models combining
with sliding window scores are used for predicting secondary
structure of a protein from its amino acid sequence. Since the
length of protein conformation segments varies in a narrow range,
we ignore the duration effect of the length distribution. The
window scores for residues are a window version of the Chou-Fasman
propensities estimated under an approximation of conditional
independency. Different window widths are examined, and the
optimal width is found to be 17. A high accuracy about 70\% is
achieved.
\end{abstract}

\leftline{PACS number(s): 87.10.+e,02.50.-r}%
\bigskip

\section{Introduction}

There is an increasing gap between knowledge of protein structure and protein
sequence. A long-term goal of the protein-folding problem is to predict the
folded 3D structure of a protein from its amino acid sequence alone. Secondary
structure prediction is an initial starting point in predicting the structure
of a protein. The first efforts to predict protein secondary structures
begun in the early 1970s (Chou and Fasman, 1974a and b). Since then, a number
of sophisticated methodologies, e.g.
statistical methods based on information theory, nearest neighbor methods,
hidden Markov models, and neural networks, have been developed. Many algorithms
examine a sequence window of 13-17 residues, assuming that the central amino
acid in the window will adopt a conformation that is determined by all the amino
acid residues in the window.

Solovyev and Shindyalov (2002) classified prediction algorithms
into three categories: methods using single sequences, combined
methods using nearest neighbors, and those using homology
information. Although the highest prediction accuracy is obtained
with methods using homology information, the methods require a
homology search of related sequences and often an additional
multiple sequence alignment. While the combined methods using
nearest neighbors directly compare the window to be predicted with
a segment database, the methods using single sequence convert such
a database to parameters, so at the prediction step no datasets
are referred, and then least computation is required. Most
window-based methods require discriminant thresholds and a
post-prediction filtering.

Hidden Markov models (HMMs) (Rabiner, 1989) have been applied to
molecular biology, in particular in gene-finding. HMMs have also
been used for protein structure studies (Asai et al.,1993; Stultz
et al., 1993; Karplus et al., 1997; Camproux et al. 1999). A
probabilistic approach similar to the gene finder Genscan has been
developed for protein secondary structure prediction without using
sliding windows (Schmidler, Liu and Brutlag, 2000). In terms of
Bayesian segmentation, the method integrated explicit models for
secondary structure classes helices, sheets and loops with other
observed structure aspects such as segment capping signals and
length distributions, and reached an accuracy comparable to the
information theory approach GOR (Garnier, Osguthorpe, and Robson,
1978; Gibrat, Garnier, and Robson, 1987; Garnier, Gibrat, and
Robson, 1996).

Compared with DNA sequences, protein sequences are generally
short, and their amino acid alphabet is of a large size 20. The
range of lengths of secondary structure segments is rather small.
The effect of duration might play a less important role.
Conformation states in most window-based predictions are single
residue conformations: coils(c), extended strands(e) and
helices(h). Correlation among residue conformations is not taken
into account. Here we develop a simple hidden Markov model with
correlation of residue conformation included for the secondary
structure prediction. We consider quintuplets of residue
conformations as new states for hidden Markov processes. We have
proposed a scoring scheme for emission probabilities based on
amino acid alphabet reduction (Zheng, 2003). We shall discuss a
scoring scheme based on sliding windows. By combining sliding
windows with HMMs, outcome from individual windows may be
synthesized, and hence error of prediction be reduced.

\section{Methods}

As in most methods, we consider 3 states $\{h, e, c\}$ generated
from the 8 states of Kabsch and Sander (1983) by the coarse-graining
$H,G,I\to h$, $E\to e$ and $X,T,S,B\to c$. Let $R\equiv R_{1:n}
=R_1R_2\ldots R_n$ be a sequence of $n$ amino acid residues, and its
corresponding secondary structure sequence be $S =S_1S_2\ldots S_n$.
The structure prediction is the mapping from $R$ to $S$.

\subsection{Hidden Markov processes}
Instead of single residue conformations $c$, $e$ and $h$, our
states of the hidden Markov chains are quintuplets of residue
conformations, or our model is of the fifth order. The main
restriction to a structure sequence is that the shortest length of
the consecutive state $h$ must be 3, and that of $e$ be 2. As a
direct outcome of this constrain, only 75 of the total $3^5=243$
possible quintuplets are legitimate. Exclusion of 7 rare ones
($eceeh$, $hceeh$, $heece$, $heech$, $heeeh$, $heehh$, and
$hheeh$) further reduces the total number of states into 68, which
are listed in Table~1. The number of transition states from any
states are at most three. The transition rates may be estimated
from hexamer and quinmer string counts. For example, the rate
\begin{equation}
T(hhccc\to hcccc) = n(hhcccc)/n(hhccc),
\end{equation}
where $n(x)$ denote the count of string $x$.

\subsection{Emission probabilities}
Emission probabilities relate residue sequences to structure sequences.
For the purpose of inferring the conformation of the central residue from
a window flanking the residue, a form of emission probabilities is
the conditional probabilities
\begin{equation}
P(R_i|W_i,\sigma_i)
\end{equation}
where $\sigma_i=S_{i-2}S_{i-1}S_iS_{i+1}S_{i+2}$, and $W_i$ is the
flanking residue window of a width $2l+1$:
$W_i=R_{i-l}R_{i-l+1}\cdots$ $R_{i+l-1}R_{i+l}$. As a convention,
we exclude $R_i$ from $W_i$. Due to the large size 20 of the amino
acid alphabet, it is impossible to directly consider residue
correlation beyond pair correlation. This means that even for
$l=1$ some approximation has to be made. We have proposed a scheme
to reduce 20 amino acids into a limited number ($\leq 4$) of
classes for coarse-graining $W_i$. Another approximation, which
has been also used in the information theory approach GOR, is the
approximation of conditional independency. The approximation
assumes
\begin{eqnarray}
P(W_i|\sigma_i) &=& \mathop{{\prod}'}_{j=-l}^{l} P(R_{j+i}|\sigma_i),\\
P(W_i|R_i,\sigma_i) &=& \mathop{{\prod}'}_{j=-l}^{l}
P(R_{j+i}|R_i,\sigma_i),
\end{eqnarray}
where the prime indicates that $j=0$ is excluded from the
multiplication. The emission probabilities are then reduced to
\begin{equation}
P(R_i|W_i,\sigma_i)=\frac{P(R_i,W_i,\sigma_i)}{P(W_i,\sigma_i)}
=\frac{P(R_i,\sigma_i)}{P(\sigma_i)}\frac{\prod'_j P(R_{j+i}|R_i,\sigma_i)}
{\prod'_j P(R_{j+i}|\sigma_i)}.
\end{equation}

Chou-Fasman propensities of amino acid residues for various conformation
is a measure of the correlation between a residue and its conformation.
The window version of the propensity may written as
\begin{equation}
\frac {P(R_i,W_i,\sigma_i)}{P(R_i,W_i)\,P(\sigma_i)}.
\label{cfw}\end{equation}
The approximation of conditional independency
\begin{eqnarray}
P(W_i|R_i) &=& \mathop{{\prod}'}_{j=-l}^{l} P(R_{j+i}|R_i),\\
P(W_i|R_i,\sigma_i) &=& \mathop{{\prod}'}_{j=-l}^{l}
P(R_{j+i}|R_i,\sigma_i),
\end{eqnarray}
leads expression (\ref{cfw}) to
\begin{equation}
\frac{P(R_i,\sigma_i)P(W_i|R_i,\sigma_i)}{P(\sigma_i)P(R_i)P(W_i|R_i)}
=\frac{P(R_i,\sigma_i)}{P(R_i)P(\sigma_i)}\frac{\prod'_j P(R_{j+i}|R_i,\sigma_i)}
{\prod'_j P(R_{j+i}|R_i)},
\end{equation}
which consists of the Chou-Fasman propensity and its window correction.

\subsection{Pseudo-counts for probability estimation}
When counts are large, probabilities can be estimated from counts.
For example, a unbiased estimate o $P(R_{j+i}|R_i,\sigma_i)$ from
counts is
\begin{equation}
P(R_{j+i}|R_i,\sigma_i)=\frac{n(R_{j+i},R_i,\sigma_i)}{n(R_i,\sigma_i)}.
\end{equation}
When counts are small, biased estimations using pseudo-counts are
commonly recommended. We use the following background pseudo-count
scheme. Assume a background probability distribution $\{\rho_i\}$.
From counts $\{n_i\}$ the estimated probabilities $\{p_i\}$ are
given by
\begin{equation}
p_i=\frac {n_i+\sqrt{n}\rho_i}{n+\sqrt{n}},\qquad n=\sum_i n_i,
\quad \sum_i \rho_i =1.
\end{equation}
In the pseudo-count estimation of $P(R_{j+i}|R_i,\sigma_i)$, the background
distribution is taken to be
\begin{equation}
\rho (R_{j+i}|R_i,S_i)=\frac{n(R_{j+i},R_i,S_i)}{n(R_i,S_i)},
\end{equation}
where, instead of quintuplet $\sigma_i$, we look at only the
central conformation $S_i$. When count $n(R_i,S_i)$ vanishes, we
simply estimate $P(R_{j+i}|R_i,\sigma_i)$ to be
$\rho(R_{j+i}|R_i,S_i)$. The background distribution for
estimation of $P(R_{j+i}|R_i)$ and $P(R_{j+i}|\sigma_i)$ are
similarly taken to be the simple amino acid frequency $\rho (R_i)$
and $\rho (R_{j+i}|S_i)$, respectively. Since $P(R_{j+i}|R_i)$ is
independent of any structure sequence $S$, no sophisticated
estimation would essentially affect the final prediction.

\section{Result}

We create a nonredundant set of 1612 non-membrane proteins for
training parameters from PDB\_SELECT (Hobohm and Sander, 1994)
with amino acid identity less than 25\% issued on 25 September of
2001. The secondary structure for these sequences are taken from
DSSP database (Kabsch and Sander, 1983). As mentioned above, the
eight states of DSSP are coarse-grained into 3 states: $h$, $e$
and $c$. This learning set contains 268031 residues with known
conformations, among which 94415 are $h$, 56510 are $e$, and
117106 are $c$. There are 296 unknown residues. The size of the
learning set is reasonable for training our parameters.

In order to assess the accuracy of our approach, we use the
following 2 test sets: Sets rs76 and casp4. A set of 124 nonhomologous
proteins is created from the representative database of Rost and
Sander (1993) by removing subunits A and B of hemagglutinin 3hmg,
which are designated as membrane protein by SCOP (Murzin et al,
1995). The 124 sequences and the learning set are not independent
of each other according to HSSP database (Dodge, Schneider and
Sander, 1998). That is, some proteins of the 124 sequences are
homologous with certain proteins in the learning set. Removing the
homologous proteins from the 124 sequences and 5 sequences with
unknown amino acid segments longer than 6, we construct Set rs76 of
76 proteins. Nonredundant 34 proteins with known structures of the
CASP4 database issued in December of 2000 are taken as Set casp4
(CASP4, 2000).

To assess prediction methods,
we calculate for each conformation the sensitivity $s_n$ and
specificity $s_p$
\begin{equation}
s_n= \frac{TP}{TP+FN},\quad s_p= \frac{TP}{TP+FP},
\end{equation}
where $TP$, $FP$ and $FN$ are site counts of the `true positive',
`false positive' and `false negative' with respect to the observed
real conformation. The total accuracy measure $Q_3$ is the total $s_n$, i.e.
the ratio of the number of correctly predicted sites to the total
number of residues to be predicted.

The model using $P(R_i|W_i,\sigma_i)$ for emission probabilities
or amino acid scores shows $Q_3$ less than 60\%. In the following,
we consider only the model using the window Chou-Fasman
propensities $P(R_i,W_i|\sigma_i) /P(R_i,W_i)$ for amino acid
scores. It is known that the length distribution of helices
deviates from a geometric distribution. The transition rate
$T(hhhhh\to hhhhh)$ may be estimated from the counts of $hhhhhh$
and $hhhhh$ or from the average helix length. By using the
conditional distribution of helix lengths with lengths not less
than $k$, the transition rate is related to the average length
${\bar l}_h$ as
\begin{equation}
T(hhhhh\to hhhhh) = 1-1/(\bar{l}_h-k+1), \qquad \bar{l}_h = \sum_{l\geq k}
l\,P(l|l\geq k).
\end{equation}
The string counts estimate the transition rate to be 0.88, while the length
distribution with $k=5$ gives 0.82. They do not coincide. We use an intermediary
value 0.83 for the transition rate. For lengths greater than 5, the approximation
of geometric distribution for lengths is good for both $c$ and $e$.

So far we have not fixed the window width. We have examined
different window widths from 9 to 19. Results are listed in Table
2. It is seen that the highest accuracy is obtained for width 17.
As full probabilistic models, the probability for individual
residue to be at various conformation may be calculated as
marginal distribution with the so-called Baum-Welch algorithm,
besides the inference of structure by searching optimal hidden
structure `path' with the so-called Viterbi algorithm. While the
inference from the Baum-Welch marginal posterior is a little
superior to that from the Viterbi algorithm in $Q_3$ value for
test set rs76, it is not the case for casp4.

Besides the popular predictor GOR IV, there is another secondary
structure predictor SSP (Solovyev and Salamov, 1991, 1994) based
on discriminant analysis using single sequence. To compare with
them, their accuracies on the same test sets are also listed in
the table.

\section{Discussion}

We have presented simple hidden Markov models to predict secondary
structure using single protein sequence. The hidden sequence is
generated by a fifth order Markov process, or by the first order
Markov process  of multi-site conformation states. Considering
that structure segments of proteins are generally short, we have
ignored the duration effect, and focused on short range
correlations.

We combine sliding window scores with hidden Markov models. Two schemes
of residue scores investigated are the conditional probability scores and
probability ratio scores of propensities. The performance of the latter
is much superior to the former. Other score schemes such as GOR scores of
information theory may also be used, and should be examined.

The fine classification of windows by quintuplet
conformation improves window scores. The HMM integrate scores of
different windows, and further reduces prediction error. The $Q_3$
accuracy for both viterbi and Baum-Welch algorithms reaches about 69.8\%.
Since the Baum-Welch algorithm provides posterior distribution over
conformation states at each position, the percentage of positions with
calculated probability for conformation higher than preset threshold
could be taken as a measure of prediction reliability.

We have treated the length distribution simply by changing the
transition rate $T(hhhhh\to hhhhh)$. A more sophisticated way to
consider the correction of length distribution is to add only a
few new states for long helices up to certain length, say 9, and
then make a geometric distribution approximation for longer
helices. The corresponding Markov graph is shown in Fig.~1. The
new states and their indices are
$$68\  hhhhh,\ 69\  hhhhh,\ 70\  hhhhhh,\ 71\  hhhhhhh,\ 72\  hhhhhhhh. $$
States 68 and 69 must follow an $h$, while state 67 follows a
non-$h$. State 69 distinguishes from 68 by being exclusively
followed by a non-$h$. However, a primary investigation shows the
improvement of this correction is insignificant.

There is scope for further improvement in our approach. We may divide
a training set into several, say 2, subsets according to residue
statistics. For this purpose, coarse-graining of amino acids would
help. The two subsets are then used separately for training to get
refined models. We may first classify a query sequence into one of the
two categories, and then apply to it the corresponding refined
model.

\begin{quotation}
{This work was supported in part by the Special Funds for Major
National Basic Research Projects and the National Natural Science
Foundation of China.}
\end{quotation}

\bigskip
\vspace{9cm}
\includegraphics{dpproh.eps}
Fig.~1 Markov graph for correction of helix length distribution.
Here, the incoming states are $18\ chhhh$ and $42\  ehhhh$, and
outgoing states are $65\ hhhhc$ and $66\ hhhhe$. Other states of
multi-$h$ are $67\  hhhhh$, $68\  hhhhh$, $69\  hhhhh$, $70\
hhhhhh$, $71\ hhhhhhh$, and $72\ hhhhhhhh$.

\newpage
Table 1. 68 quintuplet conformation states.\\

\begin{tabular}{|rc|rc|rc|rc|}
\hline
 0& ccccc& 17& chhhe& 34& eeech& 51& heecc\\
 1& cccce& 18& chhhh& 35& eeeec& 52& heeec\\
 2& cccch& 19& ecccc& 36& eeeee& 53& heeee\\
 3& cccee& 20& eccce& 37& eeeeh& 54& hhccc\\
 4& ccchh& 21& eccch& 38& eeehh& 55& hhcce\\
 5& cceec& 22& eccee& 39& eehhh& 56& hhcch\\
 6& cceee& 23& ecchh& 40& ehhhc& 57& hhcee\\
 7& cceeh& 24& eceec& 41& ehhhe& 58& hhchh\\
 8& cchhh& 25& eceee& 42& ehhhh& 59& hheec\\
 9& ceecc& 26& echhh& 43& hcccc& 60& hheee\\
10& ceece& 27& eeccc& 44& hccce& 61& hhhcc\\
11& ceech& 28& eecce& 45& hccch& 62& hhhce\\
12& ceeec& 29& eecch& 46& hccee& 63& hhhch\\
13& ceeee& 30& eecee& 47& hcchh& 64& hhhee\\
14& ceeeh& 31& eechh& 48& hceec& 65& hhhhc\\
15& ceehh& 32& eeecc& 49& hceee& 66& hhhhe\\
16& chhhc& 33& eeece& 50& hchhh& 67& hhhhh\\
\hline\end{tabular}\\

\bigskip

\bigskip Table 2. Accuracy of secondary structure predictions for
different models. Here, VI and BW stand for `Viterbi' and
`Baum-Welch' algorithms, respectively. Models discussed are indexed
with window widths.\\

\begin{tabular}{|c|c|rrrrrrr|}\hline
Model&  Test set& $S_n^c$& $S_p^c$& $S_n^e$& $S_p^e$& $S_n^h$& $S_p^h$& $Q_3$\\
\hline
9-VI & casp4& 66.08& 71.46& 63.57& 54.97& 70.37& 70.80& 67.14\\
9-BW & casp4& 66.69& 71.58& 63.29& 55.34& 70.33& 70.79& 67.32\\
9-VI & rs76 & 66.90& 71.16& 63.48& 53.43& 65.81& 68.36& 65.82\\
9-BW & rs76 & 67.53& 71.00& 63.41& 53.15& 64.76& 68.58& 65.75\\
11-VI& casp4& 67.65& 71.23& 61.37& 56.68& 71.43& 70.84& 67.72\\
15-VI &casp4& 69.51& 71.27& 63.46& 58.30& 70.98& 72.66& 68.77\\
17-VI &casp4& 70.20& 72.11& 63.80& 61.41& 73.87& 73.35& 70.21\\
19-VI &casp4& 70.29& 71.69& 62.11& 61.04& 73.97& 73.12& 69.92\\
17-BW&  casp4& 70.47& 71.53& 63.01& 61.32& 73.26& 73.22& 69.92\\
17-VI&  rs76 & 70.66& 73.26& 68.08& 57.97& 68.77& 73.35& 69.49\\
17-BW&  rs76 & 71.09& 73.05& 68.25& 58.06& 68.55& 74.13& 69.65\\
\hline
GOR4&rs76& 79.3& 66.1& 54.7& 55.3& 63.3& 68.5& 66.2\\
SSP& rs76& 59.2& 52.8& 69.0& 55.3& 67.0& 68.1& 60.0\\
GOR4& casp4& 81.9& 62.0& 43.0& 54.6& 67.1& 64.3& 63.4\\
SSP& casp4& 74.7& 58.8& 45.7& 55.6& 66.3& 63.3& 61.4\\
\hline\end{tabular}\\


\end{document}